\documentclass[reprint,eqsecnum,floats,aps,amsmath,amssymb,nofootinbib,prd,twocolumn, showpacs]{revtex4-1}

\usepackage{bm}
\usepackage{amssymb,amsmath,amsthm}
\usepackage{mathrsfs}
\usepackage{color}
\usepackage[utf8]{inputenc}
\usepackage{enumerate}
\usepackage{hyperref}
\hypersetup{colorlinks=true} 
\usepackage{appendix}
\usepackage{graphicx}
\usepackage{dcolumn}
\usepackage{bm}
\usepackage{multirow}
\usepackage{float}
\begin{document}
\title{Multi-partite analysis of average-subsystem entropies}
\author{Ana Alonso--Serrano}
\email{a.alonso.serrano@utf.mff.cuni.cz}
\affiliation{Institute of Theoretical Physics, Faculty of Mathematics and Physics, Charles University, 18000 Prague, Czech Republic}
\author{Matt Visser}
\email{matt.visser@sms.vuw.ac.nz}
\affiliation{School of Mathematics and Statistics, Victoria University of Wellington, PO Box 600, Wellington 6140, New Zealand}

\begin{abstract}

So-called average subsystem entropies are defined by first taking partial traces over some pure state to define density matrices,
then calculating the subsystem entropies, and finally averaging over the pure states to define the average subsystem entropies. These quantities are standard tools in quantum information theory, most typically applied in bipartite systems.
We shall first present some extensions to the usual bipartite analysis, (including a calculation of the average tangle, and a bound on the average concurrence),  follow this with some useful results for tripartite systems, and finally extend the discussion to arbitrary multi-partite systems.
A particularly nice feature of tri-partite and multi-partite analyses is that this framework allows one to introduce an ``environment'' for small subsystems to couple to.
\end{abstract}



\maketitle

\def\C{\mathcal{C}}
\def\tr{{\mathrm{tr}}}
\def\d{{\mathrm{d}}}
\def\g{{\mathfrak{g}}}
\def\dbar{{\mathchar'26\mkern-12mu \d}} 
\def\Hilbert{{\mathcal{H}}}
\def\H{{\mathrm{H}}}
\def\R{{\mathrm{R}}}
\def\E{{\mathrm{E}}}
\def\nn{\nonumber}
\section{Introduction}
\noindent
Entanglement in bipartite systems is completely characterized in quantum theory, but there exist very many open questions when one is seeking to quantify the entanglement among three or more subsystems; the tripartite or multi-partite situations. In this article we shall focus mainly on the study of entanglement entropy, (and mutual entropy), as defined among multiple subsystems when we impose the condition that the total global system is in a pure state. 

An extremely useful technique in this analysis is the ``average subsystem approach'' championed by Page~\cite{Page:subsystem}; whereby we shall average over all possible pure states, using this to define the average-subsystem entropies. The explicit results and bounds we have found will allow us to understand the degree of entanglement and the amount of entropy induced in making the partition into multiple subsystems.

We consider a global system partitioned into two or more subsystems, and assume the total Hilbert space factorizes as follows:  $\Hilbert=\bigotimes_{i=1}^N \Hilbert_i$. This does not mean that the states are always separable, they can be entangled among each other, or entangled with only some of the other subsystems, (provided that the Schmidt rank is greater than unity).\footnote{{Warning: For mathematicians, separable is a technical term that effectively means the Hilbert space is either finite dimensional or at worst has denumerable dimensionality. For physicists separable is a statement that a particular element of the Hilbert space can be written in factorizable form. This conflict in usage is unfortunately standard.}}    For the classification of entanglement of states in quantum information theory is usual to use the term {\it k-separable} for states of an $N$-partite system that satisfy
\begin{equation}
 \vert \Psi \rangle= \bigotimes_{i=1}^k \vert \phi_i \rangle,
\end{equation}
where $k\leq N$. The non-separable states will be called entangled. 

In order to fully quantify entanglement among different subsystems, in a $N=2$ bipartite system it is enough, (when the total system is in a pure state), to consider the von Neumann (entanglement) entropy between the two subsystems, but in the case of $N>2$ multi-partite systems this quantity does not provide us with a fully general measure of entanglement, and there is no universally agreed upon standard quantity for measuring multi-partite entanglement \cite{Guhne:2009,Walter:2017,Thapliyal:1999, Plenio:2007zz,Valido:2014}. 

Two of the quantities that have been used in the literature are, for example, the concurrence or tangle, which serve to partially quantify the entanglement in multi\-partite systems \cite{Eisert:2016,Cheng:2007}. But neither of these is really a fully satisfactory quantifiable and universal measure of entanglement. 
For instance, using the notion of partial trace to define $\rho_A= \tr_B\{ |\psi\rangle\,\langle \psi| \}$, concurrence is defined as~\cite{Cheng:2007}
\begin{equation} \label{concurrence}
\C(\rho_A) = \sqrt{2(1-\tr\{\rho_A^2\})}\,,
\end{equation}
and ``tangle'' is defined as the square of concurrence
\begin{equation} \label{tangle}
\tau=\C^2.
\end{equation}
(Concurrence and tangle are thus relatively easy to calculate.)
Is is interesting to note that concurrence and tangle are closely related to Tsallis and Renyi entropies:
\begin{eqnarray}
S_\mathrm{Tsallis} = {1-\tr(\rho^q)\over q -1 };
\qquad
S_\mathrm{Renyi} = {\ln \tr(\rho^q)\over 1-q };
\end{eqnarray}
and that the von Neumann entropy is a specific case of these entropies
\begin{eqnarray}
S_\mathrm{von\,Neumann} = \lim_{q\to1} S_\mathrm{Tsallis} = \lim_{q\to1} S_\mathrm{Renyi}.
\end{eqnarray}
It is easy to see from equations (\ref{concurrence})--(\ref{tangle}), that tangle can be expressed in terms of
\begin{eqnarray}
\tau = \C^2 = {2(1-\tr(\rho_A^2))} = 2 S_\mathrm{Tsallis}(q=2).
\end{eqnarray} 
The case $q=2$ of Tsallis entropy is often called ``quadratic entropy''. 

Other entanglement measures have been mooted, such as negativity, discord, and entanglement of formation, but they are either trivial or more difficult to work with in the average subsystem framework. 
For instance, the concept of negativity is rather subtle. One has~\cite{Cheng:2007}
\begin{equation}
\mathcal{N}(\rho) = {1\over2} \left( ||\rho^{T_A}||-1\right),
\end{equation}
involving a partial transpose with respect to the bipartite sub-system decomposition. (If the density matrix is $\rho_{ij,kl}$ then $(\rho^{T_A})_{ij,kl} = \rho_{kj,il}$. Also note that $||X||=\tr\{\sqrt{X^\dagger X}\}$.)  
Calculating the negativity can often be relatively difficult, though for pure states it simplifies to~\cite{Zyczkowski:2002,Datta}
\begin{equation}
\mathcal{N}(\rho) = {1\over2} \left([\tr(\sqrt{\rho_A})]^2 -1\right).
\end{equation}
Averaging over pure states
\begin{equation}
\langle \mathcal{N}(\rho) \rangle = {1\over2} \left(\langle[\tr(\sqrt{\rho_A})]^2\rangle -1\right).
\end{equation}
This is not trivial, but at least tractable~\cite{Zyczkowski:2002,Datta,Puchala:2016}.

Another interesting quantity is the quantum discord, although this does not directly characterize the entanglement itself, but instead measures the extent to which the correlation are quantum as opposed to classical~\cite{Brown:2013}.  However for pure states the discord reduces to the entanglement entropy, and so in the average subsystem framework we gain no extra usable information from considering the quantum discord.

Finally, another interesting measure of entanglement is given by the ``entanglement of formation'', which quantifies the minimum cost of preparing an state in terms of EPR pairs. It is given by
\begin{equation}
E_F(\rho):=\inf\left\{
\sum_{i}p_i E( |\psi_i\rangle \langle \psi_i|)  :  \rho=\sum_{i} p_i |\psi_i\rangle \langle \psi_i|
\right\},
\end{equation}
where $E(|\psi \rangle \langle \psi|)=S(\tr_B\{|\psi\rangle \langle \psi|\})$. This is also related to the concept of ``entanglement cost''.
However in the average subsystem framework the state $\rho$ is by assumption a pure state $\rho = |\psi\rangle \langle \psi|$, and the entanglement of formation therefore trivializes to the usual von~Neumann entropy $E_F(\rho)\to S(\rho_A) = S(\tr_B\{ |\psi\rangle \langle \psi|\} )$. So in the average subsystem framework we gain no extra usable information from considering the entanglement of formation.

It is also interesting to calculate the mutual information among different pairs of subsystems, or pairs of collections of subsystems, but note that this quantity is not really an entanglement measure because it merely considers the correlations between systems (that is,  the decrease of uncertainty in one subsystem when we measure the other one). What is however clear is that if the mutual information between any two subsystems is zero, then there is no entanglement between them.

\section{Average subsystem entropies}

In bipartite systems, the average subsystem entropies associated with the Hilbert space factorization $\Hilbert_{AB}=\Hilbert_A\otimes \Hilbert_B$ are defined by a simple three-step process~\cite{Page:subsystem}:
\begin{enumerate}
\item 
Take partial traces of some pure state to define two density matrices:
\begin{equation}
\rho_A = \tr_B\{ |\psi\rangle\langle \psi| \} \quad\hbox{and}\quad
\rho_B = \tr_A\{ |\psi\rangle\langle \psi| \}.
\end{equation}
\item
Calculate the two sub-system entropies (which are equal to each other): 
\begin{equation}
S_A = - \tr\{\rho_A\ln\rho_A\} = -\tr\{\rho_B\ln\rho_B\} = S_B.
\end{equation}
\item
Average uniformly over the pure states $|\psi\rangle$ to define average entropies: 

\begin{equation}
\langle S_A\rangle=\langle S_B\rangle.
\end{equation}
\end{enumerate}
By extension, in the context of a tri-partite system the obvious generalization is to consider $\Hilbert_{ABC}=\Hilbert_A\otimes \Hilbert_B\otimes \Hilbert_C$ and modify the three-step process as follows:
\begin{enumerate}
\item 
Take partial traces of some pure state to define \emph{six} density matrices: \\
\begin{eqnarray}
\rho_A &=& \tr_{BC}\{ |\psi\rangle\langle \psi| \}; \quad
\rho_B = \tr_{AC}\{ |\psi\rangle\langle \psi| \}; \quad \nonumber \\
&& \hbox{and} \quad \rho_C = \tr_{AB}\{ |\psi\rangle\langle \psi| \} ;
\end{eqnarray}
\begin{eqnarray}
\rho_{AB} &=& \tr_{C}\{ |\psi\rangle\langle \psi| \}; \quad
\rho_{BC} = \tr_{A}\{ |\psi\rangle\langle \psi| \};  \quad \nonumber\\
&& \hbox{and} \quad \rho_{CA} = \tr_{B}\{ |\psi\rangle\langle \psi| \} .
\end{eqnarray}
\item
Calculate \emph{six} sub-system entropies (three of which are pairwise equal): 
\begin{equation}
S_A = - \tr\{\rho_A\ln\rho_A\} = -\tr\{\rho_{BC}\ln\rho_{BC}\} = S_{BC}.
\end{equation}
\begin{equation}
S_B = - \tr\{\rho_B\ln\rho_B\} = -\tr\{\rho_{CA}\ln\rho_{CA}\} = S_{CA}.
\end{equation}
\begin{equation}
S_C = - \tr\{\rho_C\ln\rho_C\} = -\tr\{\rho_{AB}\ln\rho_{AB}\} = S_{AB}.
\end{equation}
\item
Average uniformly over the pure states $|\psi\rangle$ to define average entropies: 
\begin{equation}
\langle S_A\rangle=\langle S_{BC}\rangle; \quad \langle S_B\rangle=\langle S_{CA}\rangle; 
\quad \langle S_C\rangle=\langle S_{AB}\rangle.
\end{equation}
\end{enumerate}
In multi-partite systems the basic ideas are the same, but the algebra can quickly get messy; there are up to $2^N-2$ non-trivial ways of grouping the subsystems; (the empty Hilbert space, and the full Hilbert space, will be deemed ``trivial''), more on this multi-partite construction below.

\section{Bipartite entanglement}

Returning to bipartite systems, in reference~\cite{Page:subsystem} Page established a number of interesting results regarding these average subsystem entropies. In particular, even \emph{before} any averaging is enforced:
\begin{eqnarray}
 S_A = S_B \leq \ln \min\{ n_A, n_B\};
\end{eqnarray}
where $n_A = \dim(\Hilbert_A)$, and $n_B = \dim(\Hilbert_B)$. Page then considered the effect of taking a uniform average over all pure states on $\Hilbert_{AB}$. 

The central result of reference~\cite{Page:subsystem} is that the average subsystem entropy is then extremely close to its maximum possible value. (The ``average subsystem'' is very close to being ``maximally mixed''.)    
When combined with the exact result derived by Sen in reference~\cite{Sen}, wherein Sen provided a formal analytic proof of a conjecture by Page, and the discussion below,  this can be strengthened to a strict lower bound on the subsystem entropy, and in the ``thermodynamic limit'' can be strengthened to an equality.
Below we shall first present some extensions to the usual bipartite analysis, follow this with some useful results for tripartite systems, and finally extend the discussion to arbitrary multi-partite systems. 

\subsection{Exact results}

Page conjectured~\cite{Page:subsystem} and Sen proved~\cite{Sen}, (under certain mild technical assumptions, and with minor change of their notation), that when measured in natural units (nats)~\cite{nat1,nat2}, the average dimensionless entropy of a subsystem is exactly given by:
\begin{equation}
S_{n_A,n_B}  = \langle  S _A \rangle = \langle  S _B \rangle= H_{mM} - H_M - {m-1\over2M};
\end{equation}
Here
\begin{equation}
\qquad m = \min\{n_A,n_B\}, \qquad M = \max\{n_A,n_B\},
\end{equation}
where $H_n$ is the $n^{th}$ harmonic number $H_n = \sum_{i=1}^n {1\over n}$; see for instance reference~\cite{Havil}. 

\bigskip
\noindent
Writing $n=n_An_B$ for the total dimensionality, we note that $m\leq \sqrt{n}$ and $M \geq \sqrt{n}$, so that the average subsystem entropy can also be written as
\begin{equation}
S_{n_A,n_B}  = \langle  S _A \rangle = \langle  S _B \rangle= H_{n} - H_{n/m} - {m(m-1)\over2n};
\end{equation}
with $m = \min\{n_A,n_B\}\leq\sqrt{n}$.
For example
\begin{eqnarray}
&& S_{1,n_B} = 0;  \quad  S _{2,n_B} = H_{2n_B-1}-H_{n_B}; \nn \\
&& S _{3,n_B} = H_{3n_B} - H_{n_B} - {1\over n_B};
\end{eqnarray}
and for $n_B\geq 4$ we have
\begin{equation}
S_{4,n_B} = H_{4n_B} - H_{n_B} - {3\over2n_B}.
\end{equation}

\noindent
That the harmonic numbers show up here should in retrospect perhaps not be so surprising. Even in a classical statistical context  the harmonic numbers  arise in many situations where one is extremizing the Shannon entropy, $S= -\sum_{i=1}^n p_i \ln p_i$, subject to external constraints.  For example the harmonic numbers also show up in finite-space models for Zipf's law, where $p_i = {1/(i H_n)}$ and $\sum_{i=1}^n p_i = 1$~\cite{zipf}.
These harmonic numbers are well-studied in mathematics.  Note for instance the standard mathematical asymptotic result~\cite{Havil}

\begin{equation}
H_n = \gamma + \ln n +  {1\over2n} - {1\over 12 n^2} + {1\over 120 n^4} -{1\over256 n^6} + \mathcal{O}(n^{-8}).
\end{equation}
There is also an explicit bound~\cite{Havil}
\begin{equation}
H_n = \gamma + \ln n +  \epsilon_n;  \quad \hbox{where} \quad  \epsilon_n \in \left({1\over2(n+1)},{1\over2n}\right).
\label{E:havil}
\end{equation}
Here, in view of the non-overlapping nature of the bounds on $\epsilon_n$, we know that $\epsilon_n$ is monotone decreasing.
Stronger bounds are known, for instance the Franel  bound
\begin{equation}
H_n = \gamma + \ln n + {1\over2n} - {\hat \epsilon_n\over 8 n^2};  
\quad \hbox{where} \quad  
\hat\epsilon_n \in (0,1),
\end{equation}
or the result 
\begin{equation}
H_n = \gamma + \ln n +  {1\over2n} -{1\over12n^2} + {\hat \epsilon_n\over 120 n^4};  
\quad \hbox{where} \quad  
\hat \epsilon_n \in \left(0,1\right).
\end{equation}
(Yet even stronger bounds on the harmonic numbers are known, but would be overkill for current purposes.)
%
\noindent
Sometimes it is sufficient to consider the less stringent result~\cite{Havil}
\begin{equation}
H_n =  \ln n +  \tilde\epsilon_n  ;  \quad \hbox{where} \quad  
\tilde\epsilon_n \in \left({1\over n},\;1\right). 
\end{equation}
The purely mathematical explicit bound in equation (\ref{E:havil}), when combined with the exact result derived by Sen in reference~\cite{Sen},   can be strengthened to a strict bound on the bipartite entropy
\begin{equation}
S_{n_A,n_B} = \langle S_A \rangle = \langle S_B \rangle  
\in \left( \ln m - {\textstyle{1\over2}}, \; \ln m\right).
\end{equation}
That is, the average subsystem entropy is always within ${1\over2}$ nat, (less than ${1\over2\ln2} <{3\over4}$ of a bit), of its maximum possible value. Let us now formalize this statement.

\noindent
{\bf Theorem:} 
\nobreak
\begin{eqnarray}
&& S_{n_A,n_B} = \ln m + \Delta_{m,M}; \nonumber \\
&& \Delta_{m,M} 
\in \left(-{\textstyle{m\over2M}},-{\textstyle{m-1\over2M}}\right) 
\subseteq\left(-{\textstyle{1\over2}},0\right).
\end{eqnarray}
(Page says something somewhat similar in reference~\cite{Page:subsystem} but only as an estimate; this is now a rigorous bound.  There is no obvious way of \emph{cleanly} improving this bound.)

\noindent
{\bf Proof:}\\
To establish this bound, starting from the bound in equation (\ref{E:havil}), we write

\begin{eqnarray}
&& S_{n_A,n_B} = \ln m +\Delta_{m,M};  \nn \\
&& \hbox{with} \quad \Delta_{m,M}= \epsilon_{mM}-\epsilon_M - {m-1\over2M}.
\end{eqnarray}
Now since $\epsilon_n$ is monotonically decreasing we certainly have

\begin{equation}
\epsilon_{mM}-\epsilon_M - {m-1\over2M} \leq - {m-1\over2M}.
\end{equation}\
But we also have an absolute lower bound

\begin{eqnarray}
\epsilon_{mM}-\epsilon_M - {m-1\over2M} &>&  {1\over2(mM+1)} - {1\over2M}- {m-1\over2M}
 \nn \\
 &=& 
 {1\over2(mM+1)} - {m\over2M} \nn \\
 &>& - {m\over2M} > -{1\over2}. \quad
\end{eqnarray}
That is:

\begin{eqnarray}
&& S_{n_A,n_B} = \ln m + \Delta_{m,M}; \nn \\
&&\hbox{with} \qquad \Delta_{m,M} \in \left(-{\textstyle{m\over2M}},-{\textstyle{m-1\over2M}}\right)
\subset\left( -{\textstyle{1\over2}}, 0 \right).\qquad
\end{eqnarray}
This is the result we were seeking.  \hfill{$\Box$}


\subsection{Symmetric average subsystem information}

\noindent
In terms of the (symmetric) average subsystem information, (as defined by Page in reference~\cite{Page:subsystem}), we now have:
\begin{eqnarray}
I_{n_A,n_B} &=& S_{\mathrm{max};n_A,n_B} - S_{n_A,n_B}  \nonumber\\
&=& 
\ln m  - S_{n_A,n_B} 
=
- \Delta_{m,M}.
\end{eqnarray}
We then have the rigorous bounds
\begin{equation}
I_{n_A,n_B} \in \left({\textstyle{m-1\over2M}}, {\textstyle{m\over2M}}\right) \subseteq
\left({\textstyle{0}}, {\textstyle{1\over2}}\right).
\end{equation}
That is, the (symmetric) average subsystem information is always less than $1\over2$ nat; which is less than ${1\over2\ln2} < {3\over4}$ of a bit;  thus this (symmetric) definition of average subsystem information leads to a very tight bound.

\subsection{Asymmetric average subsystem information}
In contrast, in the follow-up reference~\cite{Page-curve}, Page defines (redefines) the average sub-system information in an \emph{asymmetrical} manner:
\begin{eqnarray}
\tilde I_{n_A,n_B} &=& S_{\mathrm{max};n_A} -  S_{n_A,n_B} = \ln n_A  - S_{n_A,n_B} \nn \\
&=& \ln(n_A/m) +  I_{n_A,n_B};
\end{eqnarray}
and
\begin{eqnarray}
\tilde I_{n_B,n_A}& =& S_{\mathrm{max};n_B} -  S_{n_A,n_B} = \ln n_B  -  S_{n_A,n_B} \nn \\
&=& \ln(n_B/m) +  I_{n_A,n_B}.
\end{eqnarray}
Then for the difference we always have
\begin{eqnarray}
&& \tilde I_{n_A,n_B} - \tilde I_{n_B,n_A} = \ln(n_A/n_B),
\end{eqnarray}
while for the average
\begin{eqnarray}
&& \bar  I_{n_A,n_B}  = {1\over2}\left(\tilde I_{n_A,n_B} + \tilde I_{n_B,n_A}\right)  = \ln(M) +  I_{n_A,n_B}.\qquad \nn \\
\end{eqnarray}
We note that the average $\bar  I_{n_A,n_B}\approx \ln M$ is symmetric, and, since $I_{n_A,n_B}  \in (0, {1\over2})$, it is utterly dominated by the dimensionality of the larger Hilbert space. In view of the very tight bound on $I_{n_A,n_B}$,  this means that (to within ${1\over2}$ nat) for all practical purposes we have
\begin{equation}
\tilde I_{n_A,n_B} \approx\ln\left(n_A\over m\right) = \ln\left(n_A\over\min\{n_A,n_B\}\right) ; 
\end{equation}
and
\begin{equation}
\tilde I_{n_B,n_A} \approx\ln\left(n_B\over m\right)= \ln\left(n_B\over\min\{n_A,n_B\}\right).
\end{equation}
That is, the modified average subsystem information, $\tilde I_{n_A,n_B} \neq \tilde I_{n_B,n_A}$ really says nothing much about the subsystem beyond specifying the dimensionalities of the two Hilbert sub-spaces. (Specifically, Page's asymmetric subsystem information,  $\tilde I_{n_A,n_B} \neq \tilde I_{n_B,n_A}$, contains at most ${1\over2}$ a nat of ``useful'' information.)

\subsection{Mutual information and other measures of entanglement}

It should be emphasized that mutual information is certainly not the same as what Page calls the subsystem information. (See discussion above, and references~\cite{Page:subsystem,Page-curve}, for details.) In general (using industry standard terminology) one has
\begin{equation}
I_{A:B} = S_A +S_B - S_{AB}.
\end{equation}
For the bipartite system considered by Page, where the total system is in a pure state, one has $S_A=S_B$ and $S_{AB}=0$, so yielding the particularly simple result
\begin{equation}
I_{A:B} = 2 S_A = 2 S_B.
\end{equation}
More specifically, after applying the ``average subsystem'' argument
\begin{equation}
\langle I_{A:B} \rangle= 2 \langle S_A \rangle = 2 \langle  S_B \rangle 
\approx  2\ln\min\{n_A,n_B\}.
\end{equation}
While at first glance this seems uninteresting, when combined with Page's asymmetric subsystem information this leads to
\begin{eqnarray}
&& \langle \tilde I_{A,B}\rangle   +\langle  \tilde I_{B,A} \rangle + \langle I_{A:B}\rangle \approx
\ln\left(n_A\over \min\{n_A,n_B\}\right) \nn \\
&& +\ln\left(n_B\over \min\{n_A,n_B\}\right)+ \vphantom{\bigg|} 2\ln\min\{n_A,n_B\},
\end{eqnarray}
from which we obtain the approximate sum rule
\begin{eqnarray}
&& \langle \tilde I_{A,B} \rangle  + \langle \tilde I_{B,A} \rangle + \langle \hat I_{A:B}  \rangle \approx 
\ln\left(n_A n_B\right) 
\approx \ln n_{AB}. \qquad
\end{eqnarray}
Here this approximation is now valid to within ${3\over2}$ nat. 

In counterpoint, while in bipartite systems the entanglement is well-determined in terms of the von Neumann entropy, it is possible also to bound the concurrence, and to actually calculate the average tangle.
Let $n=n_A n_B$, while $m=\min\{n_A,n_B\}$ and $M=\max\{n_A,n_B\}$, so $n=mM$. Then it is known that~\cite{Page:subsystem,Lubkin}
\begin{eqnarray}
\langle \tr(\rho_A^2) \rangle &=& {n_A+n_B\over n+1} =  {n_A+n_B\over n_An_B +1} \nn \\
&=& {m+M\over n+1} ={m+M\over mM+1}.
\end{eqnarray}
Thus the averaged tangle is given by
\begin{eqnarray}
&& \langle \tau \rangle = 2 \left(1- {m+M\over mM+1}\right) = {2(m-1)(M-1)\over mM+1}. \qquad
\end{eqnarray}
Now note that for the concurrence $\langle \C\rangle^2 \leq \langle \C^2 \rangle = \langle\tau\rangle$, so we certainly have the bound
\begin{equation}
\langle \C \rangle \leq 
\sqrt{{2(m-1)(M-1)\over mM+1}}.
\end{equation}

\subsection{Thermodynamic limit --- bipartite}
The usual classical thermodynamic limit is the infinite volume limit; and the closest one can get to this notion in a quantum von~Neumann context is to let one Hilbert space factor become arbitrarily large, while the other remains fixed.  (Specifically let $m=\min\{n_A,n_B\}$ be held fixed, while $M=\max\{n_A,n_B\} \to \infty$.) In that limit the smaller average subsystem achieves maximum entropy (maximal mixing). We can state this more formally as follows.

\bigskip
\noindent
{\bf Theorem:}
\begin{equation}
\lim_{M\to\infty} S_{n_A,n_B} =   \lim_{M\to\infty} \langle S_A \rangle =    \lim_{M\to\infty} \langle S_B \rangle  = \ln m.
\end{equation}
{\bf Proof:}\\
We note the standard mathematical result
\begin{equation}
\lim_{M\to \infty} H_{mM} - H_M = \ln m.
\end{equation}
But then
\begin{equation}
\lim_{M\to \infty}  S_{n_A,n_B} =  \lim_{M\to \infty} \left\{ H_{mM} - H_M - {m-1\over2M} \right\} = \ln m,
\end{equation}
as claimed. \hfill$\Box$\\
Thus, in the thermodynamic limit of the average subsystem approach, the finite-dimensional subsystem is always maximally entangled with the infinite-dimensional subsystem.
It is also possible to calculate the averaged tangle in this thermodynamic $M\to \infty$ limit, finding that
\begin{eqnarray}
\lim_{M\to\infty} \langle \tau \rangle = {2(m-1)\over m} = 2\left(1-{1\over m}\right).
\end{eqnarray}
Specifically, the ``tangle deficit'', the deviation from maximal tangle, is
\begin{eqnarray}
\Delta \tau &=& \lim_{M\to\infty} \langle \tau \rangle - \langle \tau \rangle
=  2\left(1-{1\over m}\right) - {2(m-1)(M-1)\over mM+1} \nn \\
&=&{2(m^2-1)\over m(mM+1)} \leq {2\over M}.\qquad\quad
\end{eqnarray}
So the ``average tangle'' is always within ${2\over M}$ nat of maximal mixing.
Similarly, in the thermodynamic limit the average concurrence is bounded by
\begin{eqnarray}
\lim_{M\to\infty} \langle \C \rangle \leq \sqrt{2\left(1-{1\over m}\right)}.
\end{eqnarray}

\subsection{Wrap up}

While the mathematical validity of the bipartite analysis is unassailable, in certain circumstances the physical relevance of the input assumptions can be questionable. In particular, the fact that the total system is always taken to be a pure state is not always entirely physically appropriate, which is one reason why we now turn to a tripartite analysis. 

We emphasize this point: Consideration of a global pure state can be very useful for some physical systems, but in some cases we cannot simply divide those systems into \emph{only} two isolated sub-systems. It might then be necessary to consider some overall encompassing environment, or to consider more than two subsystems into which the (total) Hilbert space is to be factorized; thereby
 making the system separable in that more subtle sense.

\section{Tripartite entanglement}

Let us now consider a tripartite system, to be 
modelled by the Hilbert space 
 $\Hilbert_{ABC}  = \Hilbert_A \otimes \Hilbert_B \otimes \Hilbert_C$. Let us first see how far we can get without making any averaging assumptions. Take the entire universe to be in a pure state, so at all times $S_{ABC}=0$ and the subsystem entropies satisfy
\begin{equation}
S_A = S_{BC};  \qquad S_B = S_{AC};  \qquad S_C = S_{AB}. 
\end{equation}
For the average entanglement entropy, we now have
\begin{equation}
\langle  S _A\rangle = \langle  S _{BC}\rangle \approx \ln\min\{n_A, n_B n_C \};
\end{equation}
\begin{equation}
\langle  S _B\rangle = \langle  S _{CA}\rangle \approx  \ln\min\{n_B, n_C n_A \};
\end{equation}
\begin{equation}
\langle  S _C\rangle = \langle  S _{AB}\rangle \approx \ln\min\{n_C, n_A n_B \};
\end{equation}
all three of these approximations holding to within ${1\over2}$ a nat.
This now allows us to write
\begin{eqnarray}
\langle  S _A +S_B +S_C \rangle &\approx &
\min\{ \ln n_A, \ln n_B + \ln n_C \}\nn\\ 
&& +\min\{ \ln n_B, \ln n_C + \ln n_A \} \nn \\
&& +\min\{ \ln n_C, \ln n_A + \ln n_B \}. \qquad
\end{eqnarray}
For convenience, temporarily permute $\{A,B,C\}$ so that  $n_A\leq n_B\leq n_C$. Then we have
\begin{equation}
\langle  S _A + S _B + S _C \rangle\approx  
\ln n_A+\ln n_B +\min\{ \ln n_C, \ln n_A + \ln n_B \},
\end{equation}
which implies
\begin{equation}
\langle  S _A + S _B + S _C \rangle\approx  
\ln n_A+\ln n_B +\min\{ \ln n_C, \ln n - \ln n_C \},
\end{equation}
whence we deduce
\begin{equation}
\langle  S _A + S _B + S _C \rangle\approx  
\ln n +\min\{0, \ln n - 2 \ln n_C \}.
\end{equation}
Undoing the permutation we see
\begin{eqnarray}
&&\langle  S _A + S _B + S _C \rangle\approx
\ln n \nn \\
&&+\min\{0, \ln n - 2 \max\{\ln n_A, \ln n_B, \ln n_C \} \} \quad
\end{eqnarray}
This approximation for the sum of subsystem entropies is related to the existence of bounds on variable-length compound jumps~\cite{jumps}.

\subsection{The ``rest of the universe'' --- the environment}

Now suppose one subsystem is much larger than the other two. Specifically let subsystem C denote the environment, (the ``rest of the universe'' ), while subsystems A and B are free to interact with each other, (and for now, are free to interact with the environment).  Specifically let us assume that $n_C \geq n_A n_B$.  This implies both $n_C \geq n_A$ and $n_C \geq n_B$, which furthermore implies both $n_B n_C \geq n_A$ and $n_C n_A \geq n_B$. So in this situation
\begin{eqnarray}
\langle  S _A\rangle& =& \langle  S _{BC}\rangle \approx  \ln n_A;
\quad
\langle  S _B\rangle = \langle  S _{CA}\rangle \approx  \ln n_B; \nn \\
&& \langle  S _C\rangle = \langle  S _{AB}\rangle \approx  \ln\{n_A n_B \}.
\end{eqnarray}
That is, $\langle  S _C\rangle$ is \emph{not} the total entropy of the rest of the universe, it is merely the extent to which the rest of the universe is entangled with the AB subsystem.

\subsection{Mutual information}

For the tripartite ABC system we are advocating here the situation is more interesting than for the bipartite AB system. For the tripartite system $S_{AB}=S_C$ (and $S_A\neq S_B$ in general) so
\begin{equation}
I_{A:B} = S_A +S_B - S_{AB} = S_A +S_B - S_{C}. 
\end{equation}
Now averaging over the pure states in ABC,  we have 
\begin{equation}
\langle I_{A:B} \rangle =  S_{n_A,n_B n_C} +S_{n_B,n_A n_C} - S_{n_C,n_A n_B}.
\end{equation}
So in the situation where C is a suitably large environment, $n_C \geq n_A n_B$ as discussed above, and using the harmonic numbers $H_n$ as introduced above,  we have the exact result
\begin{eqnarray}
\hspace{-10pt}
\langle I_{A:B} \rangle 
&=& \left[ H_{n_A n_B n_C} - H_{n_A n_C} - {n_A-1\over 2n_B n_C} \right] 
\nn \\
&& 
+ \left[ H_{n_A n_B n_C} - H_{n_B n_C} - {n_B-1\over 2n_B n_C} \right]
\nonumber\\
&&
- \left[ H_{n_A n_B n_C} - H_{n_C} - {n_A n_B-1\over 2n_C} \right].
\end{eqnarray}
Then after a little simplification
\begin{eqnarray}
\langle  I_{A:B} \rangle 
&=&  H_{n_A n_B n_C} + H_{n_C}  - H_{n_A n_C} - H_{n_B n_C}  
\nn \\
&&
 + { (n_A-1)(n_B-1)(n_A n_B +n_A + n_B) \over 2n_A n_B n_C}.
\nonumber
\\
\end{eqnarray}
It is now relatively easy to see  that
\begin{equation}
\langle I_{A:B} \rangle \leq {1\over2}.
\end{equation}
So in the tripartite ABC system the average mutual information between the two ``small'' subsystems A and B, never exceeds ${1\over2}$ nat. 
More formally we have the following.

\bigskip
\noindent
{\bf Theorem:} Provided $n_A n_B \leq n_C$ we have
\begin{equation}
\langle I_{A:B} \rangle \leq  { n_A n_B \over 2 n_C}  \leq {1\over2}.
\end{equation}

\noindent
{\bf Proof:}\\
We start from
\begin{eqnarray}
\label{E:xact}
\langle  I_{A:B} \rangle 
&=&  H_{n_A n_B n_C} + H_{n_C}  - H_{n_A n_C} - H_{n_B n_C}  
\nn \\
&&
 + { (n_A-1)(n_B-1)(n_A n_B +n_A + n_B) \over 2n_A n_B n_C}.
\nonumber
\\
\end{eqnarray}
and again use
\begin{equation}
H_n = \gamma + \ln n +  \epsilon_n;  \quad \hbox{where} \quad  \epsilon_n \in \left({1\over2(n+1)},{1\over2n}\right).
\end{equation}
Then the $\ln$'s and $\gamma$'s cancel and 
\begin{eqnarray}
\langle  I_{A:B} \rangle 
&=&  \epsilon_{n_A n_B n_C} + \epsilon_{n_C}  - \epsilon_{n_A n_C} - \epsilon_{n_B n_C} 
\nn \\
&&
 + { (n_A-1)(n_B-1)(n_A n_B +n_A + n_B) \over 2n_A n_B n_C}.
\nonumber\\
\end{eqnarray}
But then 
\begin{eqnarray}
\langle I_{A:B} \rangle &\leq& {1\over2 n_A n_B n_C} + {1\over 2n_C} 
\nn \\
&&
 + { (n_A-1)(n_B-1)(n_A n_B +n_A + n_B) \over 2n_A n_B n_C}. \qquad
\end{eqnarray}
That is
\begin{equation}
\langle I_{A:B} \rangle \leq 
{ n_A^2 n_B^2 - n_A^2 -n_B^2 + n_A + n_B+ 1\over 2n_A n_B n_C} .
\end{equation}
We can rewrite this as
\begin{equation}
\langle  I_{A:B} \rangle \leq 
{ n_A^2 n_B^2 \over 2n_A n_B n_C} 
- { n_A^2 + n_B^2 - n_A - n_B- 1\over 2n_A n_B n_C}, 
\end{equation}
that is
\begin{equation}
\langle  I_{A:B} \rangle \leq 
{ n_A n_B \over 2 n_C} 
- { n_A^2 + n_B^2 - n_A - n_B- 1\over 2n_A n_B n_C}, 
\end{equation}
whence
\begin{equation}
\langle I_{A:B} \rangle \leq 
{ n_A n_B \over 2 n_C} 
- { n_A(n_A-1) + n_B(n_B-1)- 1\over 2n_A n_B n_C}. 
\end{equation}
Now consider two cases:
\begin{itemize}
	\item 
	If \emph{either} $n_A >1$ \emph{or}  $n_B >1$, then we have  \mbox{$n_A(n_A-1) + n_B(n_B-1)- 1 >0$}, 
	and so certainly
	\begin{equation}
	\langle I_{A:B} \rangle \leq 
	{ n_A n_B \over 2 n_C}  \leq {1\over2}.
	\end{equation}
	\item
	If \emph{both} $n_A =1$ \emph{and}  $n_B =1$, then we go back to the exact result of equation (\ref{E:xact}) 
	and note that $\langle I_{A:B} \rangle \to H_{n_C} + H_{n_C} - H_{n_C} -H_{n_C} + 0 = 0$. 
\end{itemize}
In either case we certainly have 
\begin{equation}
\langle I_{A:B} \rangle \leq  { n_A n_B \over 2 n_C}  \leq {1\over2}.
\end{equation}
So the average mutual information between the two ``small'' subsystems A and B in the tripartite pure-state ABC system never exceeds ${1\over2}$ nat (as long as the subsystem C is dominant in the sense that  $n_A n_B \leq n_C$).~\hfill{$\Box$}

Note that the mutual information is a measure of the uncertainty remaining in one subsystem when the other one is measured, and can  be defined for both classical and quantum systems. Entanglement however is a purely quantum concept, and so it cannot be completely characterized by the mutual information. It is important to remember that in pure bipartite systems the entanglement is directly characterized by the von Neumann entropy, but not so in tripartite (multipartite) systems \cite{Facchi:2010}.

Indeed, it is not entirely clear how to formulate a specific and practical measure of the entanglement in multipartite systems that are high-dimensional (greater than dimension two). None of the quantities explained above seem fully adequate to this case, but what can certainly be said is this: The measure of global correlations (that is, the mutual entropy) is very small, so it is expected that the entanglement between systems will be also very small. We will check this specifically in the case of thermodynamic limit. 

\bigskip
\noindent
What, in counterpoint, can we say about $\langle I_{A:C} \rangle$ and $\langle I_{B:C} \rangle$, the mutual information between A or B with the ``environment'' C? We note that as long as ABC is a pure state we have (even before averaging)
\begin{eqnarray}
I_{A:C} & =& S_A +S_C - S_{AC} =  S_A +S_C - S_B;  \nn \\
I_{B:C}  &=& S_B +S_C - S_{BC} =  S_B +S_C - S_A .
\end{eqnarray}
But we already know that after averaging
\begin{equation}
\langle S_A \rangle +  \langle S_B \rangle \approx  \langle S_C \rangle,  \quad \hbox{(to within 1 nat)}.
\end{equation}
So we see
\begin{equation}
\langle I_{A:C} \rangle \approx 2 \langle  S _A\rangle; 
\qquad 
\langle I_{B:C} \rangle \approx 2 \langle  S _B\rangle,
\quad \hbox{(to within 1 nat)}.
\label{E:4.29}
\end{equation}
So \emph{these} particular mutual information scenarios do not yield any extra useful insight.

\bigskip
\noindent
We can also lump two of the sub-systems together, and  calculate the mutual bipartite information, obtaining as in the previous case that (even before averaging),
\begin{eqnarray}
I_{A:(BC)}&=&2S_A, \quad I_{B:(CA)}=2S_B, \nn \\
I_{C:(AB)}&=&2 S_C = 2S_{AB},
\end{eqnarray}
thereby verifying that (overall) we have a completely entangled system.

\subsection{Thermodynamic limit --- tripartite}

For the bipartite AB system, the whole point (usually) is to keep the total dimensionality $n_{AB}$ fixed, while letting the A and B subsystems trade dimensionality with each other. 
For the tripartite ABC system however, the environment C (the rest of the universe) is used to initially entangle the AB subsystem with the rest of the universe, but then largely ``comes along for the ride'' (as long as $n_C \geq n_A n_B$). 
So there is no real loss of generality in taking the limit $n_C\to\infty$. 
This does not mean we are making any restrictive assumptions concerning the actual thermodynamic entropy of the rest of the universe, it is a much milder statement  that the rest of the universe could in principle have an arbitrarily high dimensional Hilbert space. 
Under these conditions we have (at all times) the following limits:

\begin{equation}
\lim_{n_C\to\infty}  \langle S_A \rangle = \ln n_A;
\qquad
\lim_{n_C\to\infty}  \langle S_B \rangle = \ln n_B;
\end{equation}
which is the maximum entropy compatible with dimensionality, and 
\begin{equation}
\lim_{n_C\to\infty}  \langle  S _C \rangle = \ln (n_A n_B).
\end{equation}
which is the maximum entropy compatible with the total system being a pure state.
In this limit we therefore have the \emph{equality}

\begin{equation}
\lim_{n_C\to\infty}  \left( \vphantom{\Big{|}} \langle S_A \rangle +  \langle S_B \rangle \right)
= \lim_{n_C\to\infty}  \langle S_C \rangle,
\end{equation}
an equality which (in this limit) reproduces the classical thermodynamic arguments.  
An immediate consequence of this result is
\begin{equation}
\lim_{n_C\to\infty} \langle I_{A:B} \rangle = 0.
\end{equation}
That is, for an infinite dimensional environment C the mutual information between the subsystems A and B in a pure-state  ABC  system is zero. The fact that things simplify so nicely for an infinite dimensional environment should  perhaps not be all that surprising in view of the fact that even in purely classical thermodynamics an infinite volume limit (infinite degrees of freedom) is necessary for the existence of phase transitions. In counterpoint, an infinite dimensional environment is also necessary if for some reason one wishes to drive the Shannon entropy to infinity~\cite{shannon}.
On the other hand, it is well-known that when the mutual entropy is zero, the subsystems are completely independent, that is, we can confirm that the entanglement between these two subsystems is zero.

\section{Multi-partite entanglement}

We now seek to further generalize these considerations to explore a generic multi-partite context.
Many results carry over (with minor increase in algebraic complexity) from the bipartite and tripartite results.

\subsection{Framework}
For multi-partite decompositions the basic idea is to write
\begin{equation}
\Hilbert =  \bigotimes_{i=1}^N  \Hilbert_i;   \qquad  n_i = \dim(\Hilbert_i); \qquad n=\dim(\Hilbert),
\end{equation}
and then define partial traces
\begin{equation}
\rho_i = \tr_{{\hbox{\small (all subspaces except $\Hilbert_i$)\;}}} \rho = \tr_{\{\Hilbert/\Hilbert_i\}} \rho,
\end{equation}
or even more generally
\begin{eqnarray}
\rho_{ijk\dots}& =& \tr_{{\hbox{\small (all subspaces except $\Hilbert_i$, $\Hilbert_j$, $\Hilbert_k$, \dots)\;}}} \rho \nn \\
&=&  \tr_{\{\Hilbert/(\Hilbert_i\otimes\Hilbert_j\otimes\Hilbert_k\otimes ...)\}} \rho.
\end{eqnarray}
The Page-Sen result~\cite{Page:subsystem, Sen}  translates (at the most elementary level) to the statement that for each individual $i$ we have
\begin{equation}
 \langle  S (\rho_i) \rangle = S_{n_i,n/n_i}  = H_{m_iM_i} - H_{M_i} - {m_i-1\over2M_i}.
\end{equation}
Here in the obvious manner
\begin{equation}
\qquad m_i = \min\{n_i,n/n_i\}, \qquad M_i = \max\{n_i,n/n_i\},
\end{equation}
and where $H_n$ is the $n^{th}$ harmonic number~\cite{Havil}. 
Note we always have $m_iM_i=n$, so that we can deduce
\begin{equation}
 \langle  S (\rho_i) \rangle = S_{n_i,n/n_i}  = H_{n} - H_{n/m_i} - {m_i(m_i-1)\over2n};
\end{equation}
with $m_i = \min\{n_i,n/n_i\}$.

More generally we also have
\begin{eqnarray}
 \langle  S (\rho_{ijk\dots}) \rangle &=& S_{n_{ijk\dots},n/n_{ijk\dots}}  \nn \\
 &=& H_{m_{ijk\dots}M_{ijk\dots}} - H_{M_{ijk\dots}} - {m_{ijk\dots}-1\over2M_{ijk\dots}}, \nn \\
\end{eqnarray}
where now we define
\begin{eqnarray}
m_{ijk\dots} &=& \min\left\{n_{ijk\dots},{n\over n_{ijk\dots}}\right\}, \nn \\
M_{ijk\dots} &=& \max\left\{n_{ijk\dots},{n\over n_{ijk\dots}}\right\}.
\end{eqnarray}
Again $M_{ijk\dots}m_{ijk\dots}=n$ so that
\begin{eqnarray}
 \langle  S (\rho_{ijk\dots}) \rangle &=&S_{n_{ijk\dots},n/n_{ijk\dots}}  \nn \\
 &=& H_{n} - H_{n/m_{ijk\dots}} - {m_{ijk\dots}(m_{ijk\dots}-1)\over2n}; \nn \\
 \end{eqnarray}
where $m_{ijk\dots}$ is as defined above.

\subsection{Bounds}

One obvious comment, based on the bipartite analysis, is that for any collection $ijk...$ of subsystems with collective dimensionality $n_{ijk...}$ and $m_{ijk\dots} = \min\left\{n_{ijk\dots},{n/ n_{ijk\dots}}\right\}$ we have the rigorous bound 
\begin{eqnarray}
&& \langle  S (\rho_{ijk\dots}) \rangle = S_{n_{ijk\dots},n/n_{ijk\dots}}  = \ln(m_{ijk\dots}) + \Delta; \nn \\
&& \Delta \in \left(-{1\over2},0\right).
\end{eqnarray}
So we see that any collection of these average subsystems is close to being maximally mixed, in fact within ${1\over2}$ a nat of maximal mixing.

\subsection{Mutual information}

The mutual information is in principle easy to deal with, just algebraically messy. \\
Let us define three disjoint subsystems ABC as follows:
\begin{eqnarray}
A: \Hilbert_{ijk...} &=&  (\Hilbert_i\otimes\Hilbert_j\otimes\Hilbert_k\otimes ...); \nn \\
B: \Hilbert_{pqr...} &=&  (\Hilbert_p\otimes\Hilbert_q\otimes\Hilbert_r\otimes ...);
\end{eqnarray}
where none of the indices overlap, and then define C by setting
\begin{equation}
C:  \Hilbert_{({ijk...pqr...})^*} = \Hilbert/( \Hilbert_{ijk...} \otimes  \Hilbert_{pqr...} ).
\end{equation}
Then we can immediately apply the tripartite analysis to this situation.
In particular as long as A and B are ``small'' and C is ``large'' for the mutual information we certainly have 
\begin{eqnarray}
\langle I_{A:B} \rangle &=& \langle I_{(ijk...):(pqr...)} \rangle \leq 
 { n_{ijk...} n_{pqr...} \over 2 n_{({ijk...pqr...})^*}}   \nn \\
 &=&
 { n_{ijk...}^2 n_{pqr...}^2 \over 2 n}  \leq {1\over2}.
\end{eqnarray}
So the average mutual information between any two ``small'' collections of subsystems, A and B, in the multi-partite pure-state system never exceeds ${1\over2}$ nat as long as the subsystem collection C is dominant in the sense that $ n_{ijk...} n_{jkl...} \leq n_{({ijk...pqr...})^*}$. 
More prosaically by ``small'' collections we mean that when considering the collections $A$, $B$ and $C$, the product  of the two total dimensions of the ``small'' collections is less than or equal to the dimension of whatever is left over: $n_A n_B 
\leq  n_C =  n_{total} / ( n_A n_B )$. That is, for ``small'' collections we require $n_A^2 n_B^2 \leq n_{total}$. As promised the only slightly tricky thing is keeping track of all the indices. 

Note that this result is only applicable to pairs \footnote{In Ref. \cite{Borras:2009} (and references therein)  an interesting alternative approach to multipartite entangled systems was developed, analyzing bipartite measures for pure states averaged over all possible bipartitions of the system. This was done in order to characterize the entanglement of the system by a single measure, with a view to then studying the robustness of entanglement. This approach seems orthogonal to the average subsystem approach discussed herein.} of ``small'' collections of subsystems; pairs of ``large'' collections, (and also any ``small'' collection when compared with a ``large'' collection), will have some degree of entanglement, often close to maximal. Overall, the system is not separable at all.

\subsection{Thermodynamic limit --- multi-partite}

In the thermodynamic limit, letting $n\to\infty$ while keeping $n_{ijk..}$ fixed, we  see
\begin{equation}
\lim_{n\to\infty} \langle  S (\rho_{ijk\dots}) \rangle = \lim_{n\to \infty} S_{n_{ijk\dots},n/n_{ijk\dots}}  = \ln(n_{ijk\dots}).
\end{equation}
So we see that, in the thermodynamic limit, any collection of these average subsystems is exactly maximally mixed, and maximally entangled with the complementary collection of average subsystems, (not just close to maximal). Indeed in the thermodynamic limit the mutual information between any collection of (non-overlapping) average subsystems vanishes:
\begin{equation}
\lim_{n\to\infty} \langle I_{A:B} \rangle = \lim_{n\to\infty} \langle I_{(ijk...):(pqr...)} \rangle = 0.
\end{equation}
As in the tripartite case, we can affirm that the entanglement between pairs of ``small'' collections of subsystems is zero in the thermodynamic limit. That is, we can always choose a pair of ``small'' collections of subsystems of the overall entangled system that will be completely unentangled.

\section{Discussion}

In this article we have studied so-called ``average subsystem'' entropies~\cite{Page:subsystem,Datta,Sen,Page-curve,Lubkin} in bipartite, tripartite, and multi-partite scenarios. This analysis takes the ``universe'' to be in a random pure state, splits the universe into sub-systems which are \emph{not} pure, (and so have non-zero subsystem entropies),
and averages over the pure states.  This model for subsystem entropies has been found to be a useful one in many different contexts. 
The most common applications found in the literature are for bipartite systems~\cite{Page:subsystem,Datta,Sen,Page-curve,Lubkin}, where subsystem entropies are typically within 1 nat of maximal mixing, but we have argued herein that it is often more useful to look at tri-partite or even multi-partite decompositions of the universe. (Indeed multi-partite decompositions have attracted and continue to attract considerable attention~\cite{Walter:2017,Thapliyal:1999,Valido:2014,Brown:2013,Facchi:2010,Luo:2016,Hwang:2016otg}.)

Tri-partite analyses are particularly useful in that they allow one to introduce a notion of ``environment'' for the other two subsystems to interact with.
In the tri-partite context the situation is cleanest when two subsystems are ``small'' compared to the third  (specifically, $n_A n_B \leq n_C$). In this situation the subsystems A and B (and even AB) are close to maximally entangled with C, while the entanglement between A and B (as measured by average mutual information) is utterly minimal (less than 1 nat). The ``thermodynamic limit'' ($n_C\to\infty$, while $n_A n_B$ is held fixed) is particularly well behaved, with the subsystems A and B (and even AB) maximally entangled with C, while the entanglement between A and B (as measured by average mutual information) is zero. 

In the multi-partite context it is convenient to lump the individual subsystems into ``collections'' A, B, C. As long as two collections are ``small'' compared to the third, then the tri-partite analysis sketched above will still hold, including the existence of the thermodynamic limit.

While we were originally inspired to consider these ideas based on an analysis of the entropy budget in the Hawking radiation process~\cite{Alonso-Serrano:2015,Alonso-Serrano:2017,burning}, and also in a cosmological context~\cite{ana1,ana2}, wherein it can be seen that the quantum consideration of the rest of the universe can be crucial, absolutely nothing in the current article specifically depends on the physics of black holes or any cosmological model, (neither general relativity black holes nor analogue black holes~\cite{Unruh:1981,Visser:1993,Visser:1998,Visser:1998b,Visser:2001a,LRR,Barcelo:2006,Visser:2007,Visser:2010,Lake-Como}; neither experimental~\cite{Weinfurtner:2010,Weinfurtner:2013,Unruh:amplifier,Unruh:measured,Steinhauer:2014,Steinhauer:2015a,Steinhauer:2015b,Belgiorno:2010,Schutzhold:2010,Belgiorno:2010b,Rubino:2011,Liberati:2011} nor theoretical~\cite{Visser:1997,Barcelo:2004,trapped,Visser:2001,Barcelo:2010xx,Barcelo:2010yy,thermality,sparsity,Visser:1992}), and nothing herein depends on any specific aspect of the Hawking evaporation process --- such as the distinction between event and apparent horizons~\cite{dublin,weather,Hawking:2015,observability}. The use of bipartite, tripartite, and multi-partite decompositions of Hilbert spaces, and the use of the average subsystem approach, are general tools of quantum information theory, and we have tried to carefully separate out the general features from the specific applications~\cite{Page:subsystem,Guhne:2009,Walter:2017,Thapliyal:1999, Plenio:2007zz,Valido:2014,Eisert:2016,Cheng:2007,Datta,Brown:2013,Sen}. Making this clean conceptual separation has allowed us to completely side-step the highly contentious issues associated with the ``information puzzle''~\cite{info-puzzle,firewall,apologia,Chen:2015,Mathur:2015,Nomura:2012a,Nomura:2012b,Israel:2014,Albrecht:2014,w/o-loss,Mathur:2009,Preskill:1992,Ashtekar:2005,Hayward:2005,Hayward:2005b,Hayward:2005c,Braunstein:2009my,Chakraborty:2017pmn,Bardeen:2014,Luo:2016}.
Of course there are implications for the ``information puzzle'' --- see specifically references~\cite{Alonso-Serrano:2015,Alonso-Serrano:2017} --- but we shall not touch on these issues in the present article.

\acknowledgments{
The authors are grateful to Eduardo Mart\'{i}n-Mart\'{i}nez for very useful conversations. AA-S is supported by the grant GACR-14-37086G of the Czech Science Foundation. MV is supported by the Marsden fund, administered by the Royal Society of New~Zealand.}

 

\begin{thebibliography}{99.}
\bibitem{Steinhauer:2015b}
J.~Steinhauer,
``Observation of thermal Hawking radiation and its entanglement in an analogue black hole'',
Nature Physics, doi:10.1038/nphys3863
\end{thebibliography}
\end{document}